# Foundation Models and Information Retrieval in Digital Pathology

H.R. Tizhoosh

Kimia Lab, Mayo Clinic, Rochester, MN, USA

# Introduction

The surge in adoption of digital pathology has the potential to revolutionize medical diagnosis by allowing computerized analysis of tissue images (Pantanowitz 2010; Aljanabi 2012; Hanna2020). Central to this technology is the digitization of formalin-fixed, paraffin-embedded (FFPE) tissue sections mounted on glass slides. This process converts physical tissue samples into high-resolution, gigapixel digital images called *whole slide images* (WSIs) (Kumar2020; Evans2022). These WSI files contain detailed patterns of tissue morphology, enabling the application of computer-vision algorithms in diagnostic pathology. Pathologists can now analyze tissue images seamlessly on computer screens at various magnifications (Griffin2017). This shift from light microscopes to digital displays allows for easier visual inspection of anatomic clues that may indicate specific diseases. Additionally, and in contrast to physical sharing of glass slides, digital pathology facilitates fast and efficient sharing of tissue images for consultation and annotation, as well as quantitative analysis (Kiran2023; Parwani2020).

In recent years, the advent of deep learning and steady and rapid progress in artificial intelligence (AI) have been incessantly pushing forward the research in digital pathology. The availability of a myriad of AI models, among others, has enabled the application of many sophisticated operations in computational pathology, operations such as tissue segmentation, tumor detection and other assistive actions for diagnosis (Tizhoosh2018; Klein2021; Bera2019). In addition, digital pathology is offering new frameworks for consultations (i.e., securing second opinions) for complex and rare cases, and facilitating next-generation telemedicine (Zarella2023; Browning2021). Although in many cases, AI models are still unable to deliver acceptable diagnostic accuracy and efficiency for clinical utility, the perspectives of enhanced collaborations, efficient computer-aided analysis, and the development of a large set of quantitative methods are depicting a rather optimistic view of the future in histopathology (Niazi2019). However, as one of the major obstacles, the management of sheer volume and complexity of WSI data present a daunting challenge to the digital pathology community. Efficiently assembling well-organized (small or large) repositories of tissue images for information access is a critical task for the future of computational pathology (Kalra2020b; Tizhoosh2021; Hanna2022); information retrieval is experiencing a renaissance in light of all these developments.

# Information Retrieval

For decades, information retrieval (IR) has played a crucial role in structuring, identifying, and facilitating user-friendly access to information within complex clinical datasets (Singhal2001; Manning2009). IR systems are designed to efficiently match user queries (predominantly in form of 'text') with massive amounts of data to locate and retrieve the desired information (Guo et al., 2016; Sivarajkumar, 2024). The most fundamental and evident application of IR is searching in large archives of medical literature. Platforms like PubMed and Google Scholar are essential tools for researchers, clinicians, and even patients seeking information in research papers, clinical trials, and publications (Hersh2008; Vanopstal2013; Ting2013). These platforms leverage IR technologies to identify relevant articles based on keywords, medical terminology, and other search criteria.

Perhaps the most common usage of IR systems involves searching Electronic Health Records (EHRs). IR functionalities allow us to efficiently search for and retrieve patient information from large hospital archives (Wang 2019; Reis2016). Timely and accurate diagnosis, effective and individualized treatment planning, and productive and timely research, all require access to patients' medical history and clinical data (McInerney2020). Search and retrieval, therefore, connect the more general category of decision support systems, a set of software tools to provide clinicians with relevant patient information at the point of care. By locating patient data in the context of relevant literature, identifying personalized treatment options, and predicting patient outcomes, IR can be a major contributor for paving the way for individualized medicine (Kamath2021; Sivarajkumar2024). Finding and retrieving patients' unique characteristics in their medical history can then be used to customize the treatment plan. In context of research, drug discovery and development is another key application of IR systems. By searching vast and heterogenous repositories of chemical compounds and biological information, search and retrieval can identify potential drug candidates and most promising development trajectories, accelerating the complex drug discovery process (Chaudhary2016; Laurianne2020).

While the significant role of present and future IR systems in medical advancements is rather obvious, the implementation of sophisticated IR platforms has been experiencing formidable roadblocks. The rapid growth of multimodal patient data makes veracity, accuracy, and quality of big data a daunting challenge. Additionally, investigating user-friendly interfaces for complex IR queries, as an academically less alluring task, has not received enough attention. In the realm of digital pathology, with its emphasis on whole slide imaging, and with existence of multiple H&E and IHC whole slide images for each patient, *image search* as a specific form of information retrieval has recently gained more traction among researchers.

## Image Search

Human vision is central to information gathering, processing and analysis in most adults. This is especially true in medicine, where radiology and pathology images serve as vital diagnostic tools [Cornsweet2012; Suetens2017]. In pathology, the gold standard for diagnosing many diseases, particularly cancers (neoplasia), relies heavily on the visual inspection of tissue samples, whether digitally on a computer screen or analogously through the eyepiece of a light microscope [Rosai2007; Tseng2023]. Image search, as a form of visual information retrieval, offers a powerful new approach by bridging the gap between words, molecular data, and the visual patterns observed in tissue samples [Tizhoosh2021]. This technology, sometimes referred to as "reverse image search," allows pathologists to use IR systems and submit a "*tissue image as a query*" (and via typing words) and retrieve tissue samples with similar morphological features. This specific type of image search in an atlas (an indexed archive) is called Content-Based Image Retrieval (CBIR) (see Figure 1), and it holds significant potential for various applications in histology, histopathology, and cytopathology [Zhou2008; Babaie2017; Pantanowitz2021; Tayebi2022].

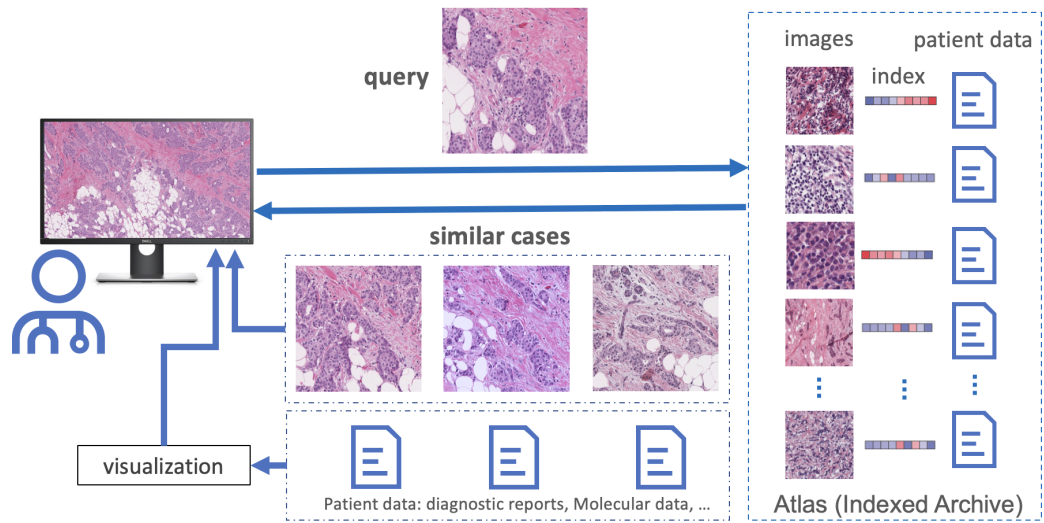

*Figure 1.* CBIR in digital pathology. Pathologists can submit a query (a 'patch' or a WSI) to an atlas (indexed archive of tissue images, patches or WSIs). The atlas generally contains patient data or links to other archives to request metadata and multimodal patient data. Similar cases can be retrieved after search for query tissue and sent back to the pathologist's workstation for visualization.

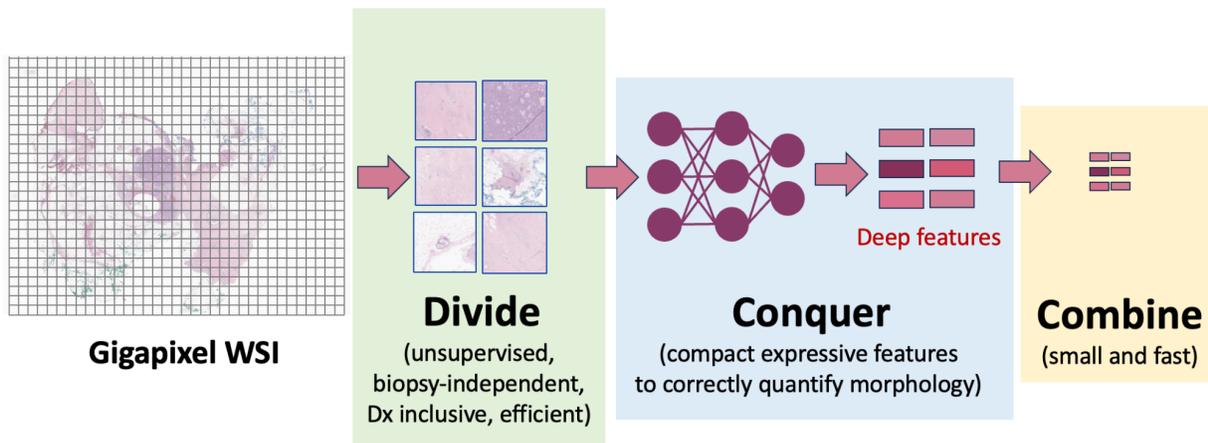

*Figure 2.* Processing WSIs with their gigapixel dimensionality requires meticulous design of the "divide & conquer" concept. The <u>divide</u>, splitting the WSI into a small number of patches, has to be unsupervised, biopsy-independent, diagnostically inclusive, and efficient. The <u>conquer</u> is mostly performed by feature extraction from a properly designed deep network. A last stage of <u>combine</u> may use some notion of compression or encoding for fast processing and lean storage of the features [Tizhoosh2024].

Due to their gigapixel size, complex nature and lack of labeled data, processing WSIs for search and retrieval purposes requires a *divide-and-conquer* approach (Figure 2). This strategy allows for the construction of an "atlas" for a specific anatomical site (lung, breast, prostate, skin etc.) and corresponding diagnoses (Figure 3). The word "atlas" has been in use for centuries and generally means "*a collection of maps or charts, usually bound together… In addition to maps and charts, atlases often contain pictures, tabular data, facts about areas, and indexes of placenames keyed to coordinates of latitude and longitude or to a locational grid with numbers and letters along the sides of maps.*" (Encyclopaedia Britannica). In context of the information retrieval, the word atlas can be understood as follows:

> An **atlas** *for a specific disease is a structured and indexed collection of patient data, well-curated to represent the spectrum of the disease diversity. To each patient entry*

*in the atlas an outcome may be attached including but not limited to primary diagnosis, and successful treatment. The outcome is ideally free from variability. The <u>indexing</u> is a computerized process to create and use the atlas mainly consisting of patient data representations that can be easily and efficiently searched and matched to locate semantically, biologically, anatomically, clinically and genetically correct pattern similarities.*

By facilitating the analysis and management of high-volume histological images through deep features for indexing (i.e., patient representation) and atlas creation, CBIR can significantly assist pathologists in their daily workflow. One application of CBIR in digital pathology involves quality assurance. For instance, CBIR can help identify potential discrepancies between a patient's tissue image and similar cases within the archive. More broadly, CBIR contributes to disease diagnosis by enabling the comparison of a query image with well-characterized cases in an atlas of WSIs with known diagnoses (Kalra2020a; Kalra2020b). This retrieval of visually similar WSIs with confirmed diagnoses empowers pathologists with evidence-based information, allowing them to identify disease patterns with greater confidence (Tizhoosh2021). The same atlas-building principles can be applied to develop evidence-based platforms for advanced triaging and treatment planning tools.

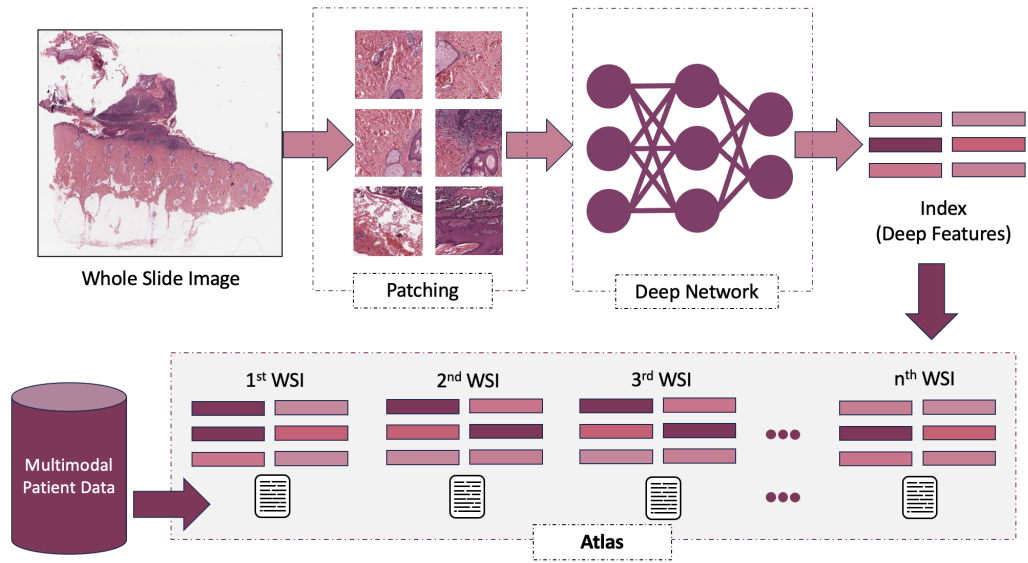

*Figure 3.* An "atlas" for CBIR in histopathology can be constructed by indexing WSIs and storing indexes with links to other patient data. It is paramount that index is compact and does not need excessive storage overhead.

In recent years, several CBIR solutions have been proposed for histopathology. Among others, hashing-based image retrieval [Zhang2014] (short HBIR), visual dictionary (or bag of visual words, BoVW, for instance [Zhu2018]), SMILY [Hegde2019], Yottixel [Kalra2020a], SISH [Chen2022, Sikaroudi2023], and RetCCL [Wang2023]. Recent examples also include HSDH (histopathology Siamese deep hashing) [Alizadeh2023] and High-Order Correlation-Guided Self-Supervised Hashing-Encoding Retrieval (HSHR) [Li2023].

Table 1 summarizes various CBIR methods for histopathology. While some, like BoVW, Yottixel, and SISH, function as complete search engines, others focus on specific aspects of information retrieval. For example, HBIR, SMILY, and HSHR primarily deal with image patches rather than entire WSIs. Additionally, HSDH and RetCCL are centered on specific deep learning models they propose. A critical challenge in histopathology search is dividing WSIs into manageable sub-images (i.e., tiles or patches) for efficient and inclusive retrieval, independent of the original biopsy

(Tizhoosh2024). This key step (see Figure 2) is often overlooked by CBIR solutions by prioritizing speed or accuracy over storage. SISH exemplifies this approach, boasting constant search time but demanding excessive storage requirements. In contrast, Yottixel's patching method, which creates a "mosaic" of image patches, has become a foundation for other CBIR approaches due to its effectiveness (Kalra2020a). This method, combined with deep feature barcoding, enables both fast search and efficient storage (Figure 4). We can generally postulate the following general rules:

- Image search will be very slow if no patch selection is applied.
- Image search will not be accurate if deep features emanate from a sub-optimally trained model.
- Image search will be very slow if indexing is complicated.
- Image search will be infeasible/impractical if its indexing requires excessive storage.

*Table 1.* Overview of CBIR solutions for histopathology (also see [Tizhoosh2024, Lahr2024]).

|  | Patching | Features | Encoding | Searching | WSI Matching | Speed | Memory Need |
|---|---|---|---|---|---|---|---|
| **HBIR** | none | none | Hashing | Hashcode matching | No | Fast | Low |
| **BoVW** | Visual words | Any feature extraction method | Counting | Histogram matching | Yes | Fast | Very low |
| **SMILY** | none | Custom | none | Feature matching | No | Slow | Very High |
| **Yottixel** | mosaic | Any feature extraction method | Barcoding | Barcode matching | Yes | Fast | Very low |
| **SISH** | Yottixel's mosaic | DenesNet | Yottixel's barcoding | Barcode matching | No | Sluggish | Excessive |
|  |  | Custom Autoencoder |  | Tree matching |  |  |  |
| **RetCCL** | Yottixel's mosaic | Custom | none | Feature matching | No | Slow | Very high |
| **HSDH** | none | Custom | none | Feature matching | No | Slow | Very High |
| **HSHR** | none | Custom | Hypergraphs | Correlation fusion | Yes | Slow | Very High |

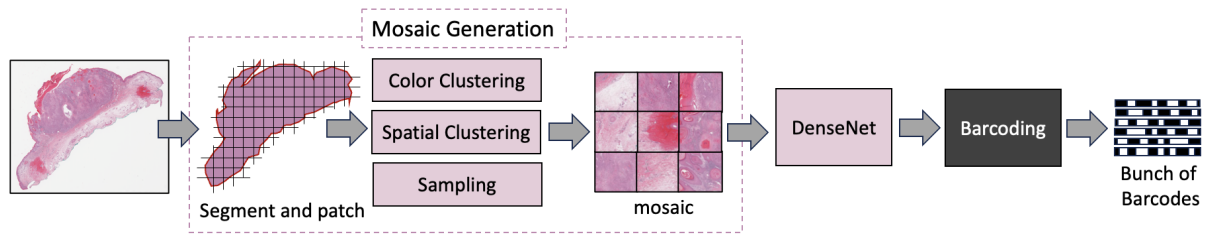

*Figure 4.* The indexing structure of Yottixel [Kalra2020a] has sparked inspiration for other search methodologies [Sikaroudi2023, Lahr2024]. The mosaic and barcode stages contribute significantly to the rapid and efficient patch/WSI search process.

## Validation of Image Search Methods

Lahr et al. recently conducted a thorough analysis and validation of BoVW, Yottixel, SISH, and RetCCL, utilizing both internal and external data [Lahr2024]. Several other methods, namely SMILY, HSDH, and HBIR, were excluded due to a lack of a patching algorithm, while HSHR was excluded due to the unavailability of code (the absence of a patch selection could be an additional reason for ignoring a search scheme). The findings suggest that BoVW and Yottixel present comprehensive search solutions, combining high speed and efficient storage, offering valuable capabilities. However, to achieve high accuracy, integration with a well-trained backbone network and adjustments to the primary site are essential.

As Yottixel is a commercial product, BoVW provides researchers with more freedom to explore various improvement avenues (also, SISH provides limited access for research and usage because it borrows Yottixel's patented barcoding [Tizhoosh2022]). The manual settings in Yottixel's mosaic, particularly regarding cluster number and sampling percentage, could benefit from automation. Additionally, a logarithmic barcode comparison approach might enhance the speed of median-of-minimum Hamming distance calculations in Yottixel. On the other hand, Lahr et al. express concerns about SISH, which, as a Yottixel variant, deviates from Occam's Razor principle, introducing speed and scalability challenges due to its unnecessarily complex structure. SISH's reliance on vEB trees (an obsolete data structure mainly suitable for priority queues and not capable of dealing with hyperdimensional domains such as WSI analysis) with exponential space requirements renders it impractical for large datasets, impeding the loading and processing of terabytes of data. Regarding RetCCL, despite being labeled a search engine, it primarily focuses on the CCL network and lacks expressive embeddings for tissue morphology. Consequently, it struggles to qualify as an effective search engine.

In their conclusion, Lahr et al. provide, for the first time for histopathology, a ranking scheme for search methods, which can be found in Table 2 for investigated variations of four image search methods. As storage is a crucial factor in digital pathology, the storage overhead of search engines (additional storage they need to save indexing records) becomes a pivotal factor for search strategy selection (see Table 3).

# Large Deep Models

The design and development of large deep models started with Large Language Models (LLMs) [Yang2023, Zhao2023]. They are built on deep learning topologies, particularly <u>transformer</u> architectures [Liu2024], and are disrupting the way we deal with natural language processing. LLMs are trained with colossal amounts of diverse text documents to generate human-like language. LLMs like GPT-3 (Generative Pre-trained Transformer 3) with an astronomical 175 billion parameters (i.e., weights similar to synapses in the human brain) are among the largest language models created to date [Wu2023]. LLMs are generally trained using many massive cohorts of text documents to learn general language patterns. Afterwards they may be additionally fine-tuned to become specialized for specific downstream tasks [Zhang2023].

The nuances and intricacies of natural language is learnt first by an LLM to later adapt to different contexts. LLMs can summarize documents, translate a document into different languages, and answer questions in an intuitive way [Hadi2023, Laban2023]. They can create content, even generate ideas, and writing different creative and professional text formats like poems, musical pieces, and computer programs. They can also serve as conversation chatbot to perform all these tasks in a user-friendly manner [Freire2024]. Despite the fascination and obvious utilities, LLMs also exhibit major pitfalls. They may <u>hallucinate</u> and generate wrong text [Rawte2023, Xu2024]. They may also - due to relying on the colossal data from public sphere - be biased [Salewski2024]. Ethical considerations about data privacy is another issue of LLMs [Peris2023]. The pathology community is concerned about these challenges [Ullah2024, Hart2023]. However, the potentials of LLMs to derive clinical factors for extraction of relevant information from huge medical records, among others pathology reports [Choi2023]. Other researchers and clinicians point to the expected future of LLMs and encourage pathologists to "embrace this technology" by identifying in what ways LLMs may support pathology for educational, clinical, and research purposes [Arvisais2024]. As well, active involvement of pathologists is necessary in improvement of AI and in helping to "design user-friendly interfaces" to integrate AI within the pathology workflow.

*Table 2. Lahr et al. ranked the performance of search engines based on their F1-score (top-1, majority of top-3, and majority of top-5), indexing time, searching time, failures, and storage (total performance ranking between 1 (best) and 6 (worst)) [Lahr2024].*

|  | Top-1 | MV@3 | MV@5 | Indexing time | Searching time | Failures | Storage | Total Ranking |
|---|---|---|---|---|---|---|---|---|
| Yottixel* | 3 | 2 | 1 | 2 | 1 | 1 | 2 | **1.71** |
| Yottixel-KR[1] | 2 | 1 | 2 | 2 | 1 | 2 | 2 | **1.71** |
| Yottixel-K[2] | 1 | 3 | 4 | 2 | 1 | 2 | 2 | **2.14** |
| BoVW | 6 | 6 | 6 | 1 | 2 | 2 | 1 | **3.43** |
| SISH* | 4 | 4 | 3 | 3 | 3 | 4 | 4 | **3.57** |
| SISH-N[3] | 5 | 5 | 5 | 3 | 3 | 4 | 4 | **4.14** |
| RetCCL-N[3] | 8 | 8 | 7 | 4 | 4 | 3 | 3 | **5.28** |
| RetCCL* | 7 | 7 | 8 | 4 | 4 | 3 | 3 | **5.43** |

\* As originally proposed
[1] DenseNet replaced by KimiaNet, and using ranking after search, [2] DenseNet replaced by KimiaNet
[3] SISH with no ranking after search, [4] RetCCL with no ranking after search

*Table 3. Index storage requirements for different search methods based on recent validation studios [Lahr2024]. BoVW is the most storage efficient approach whereas SISH requires 31 terabytes indexing storage for one million WSIs making it impractical for large datasets. At the present level of technology, loading terabytes of indexing data into memory is simply impossible.*

|  | **Index size per WSI (Kilobyte)** | **Index size for one million WSIs (Gigabyte)** |
|---|---|---|
| **BoVW** | 0.03 | 10 |
| **Yottixel** | 0.38 | 119 |
| **RetCCL** | 5.76 | 1,800 |
| **SISH** | 97.50 | 31,000 |

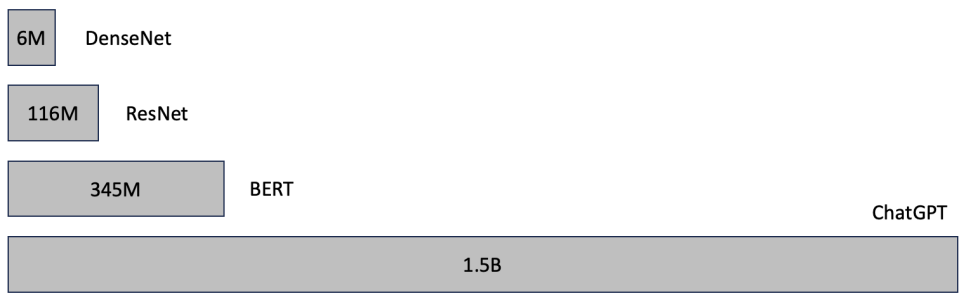

*Figure 5. Size of CNNs like DenseNet and ResNet versus LLMs like BERT (one of the smallest LLMs) and ChatGPT. LLMs like LlaMa are with 65B parameters and PaLM with 540B parameters are much larger than ChatGPT.*

Perhaps more significant for histopathology, large vision-language models (LVLMs) are another type of large deep network constellation with the capabilities of LLMs but also fluent in computer vision, hence being able to process tissue images. The most notable model in this category is CLIP, short for Contrastive Language-Image Pre-Training [Radford2021]. Relying on bidirectional training between images and captions, LVLMs can generate both human language and digital images. Expectedly, such models are trained on massive datasets comprising images and corresponding textual descriptions, e.g., captions. The training on image-text pairs enables LVLMs like CLIP to build a two-way bridge between digital images and textual descriptions. In the case of CLIP, that was 400,000,000 image-text pairs. With such colossal training data, LVLMs become capable of answering questions about the content of images. Equipped with this additional capability compared to LLMs, LVLMs can perform image classification, object detection, and image retrieval based on textual descriptions. The *conventional* image search can be made tremendously more useful by LVLMs by offering the possibility to the user to utilize natural language queries for a more intuitive and efficient experience but retrieving content-based image matches.

Like LLMs, LVLMs also face challenges. Among others, they need a very large set of image-text pair of high quality, facing biases in training data, restricted understanding of complex visual concepts such as tissue images of rare diseases, and ethical concerns regarding generated content. Cross-modal topologies are the transition from conventional deep networks toward LVLMs. ***LILE (Look In-Depth before Looking Elsewhere)*** [Maleki2022] is an example for such architectures, a dual attention network using transformers, that takes images and texts as inputs and extracts feature representations for each of those using a dedicated transformer. Then, a self-attention module [Vaswani2017] is applied. LILE is an architecture with a new loss term to help represent images and texts in a joint latent enabling bidirectional retrieval (Figure 6). LILE was trained and tested with ARCH [Gamper2021], a dataset of more than 7,500 image-text histopathology images and showed better performance than foundation models like CLIP (see next section).

Cross-modal retrieval generally aims to identify a joint latent (feature) space where different modalities, such as image-text pairs, can be brought into close alignment. However, the main challenge often lies in the representation of expressive features for tissue morphology. LLMs can generally be trained relatively easily, whereas visual models often struggle due to the scarcity of labeled data [Maleki2024]. To address this issue, a novel concept called "*harmonization*" has been introduced into processing histopathology images that can improve DINO (distillation without supervision) [Caron2021]. The harmonization of scale, a paramount factor in digital pathology, refines the DINO paradigm through a novel patching approach, overcoming the complexities posed by gigapixel whole slide images (see Figure 7).

# Foundation Models

Foundation models (FMs) [Bommasani2021] mark a recent milestone in artificial intelligence that could drastically transform the role of machines within society and medicine. Generally, FMs are understood to be "base models" that offer the adaptability to different domains. As such, one could understand FMs as very large models with unique capability to adjust to downstream tasks [Touvron2023]. By training on colossal amounts of data (which understandably requires massive computation power), FMs can learn pervasive patterns and general relationships within a enormous amount of unlabeled data, and not for specific tasks. Of course, FMs must be trained with good quality, conflict-free and diverse data types such as text and images (e.g., duplicate and new-duplicate data instances must be filtered out from the training data). As the data cannot be labeled at such massive scale that FMs require, the dominant training paradigm is unsupervised or self-supervised [Zhao2023].

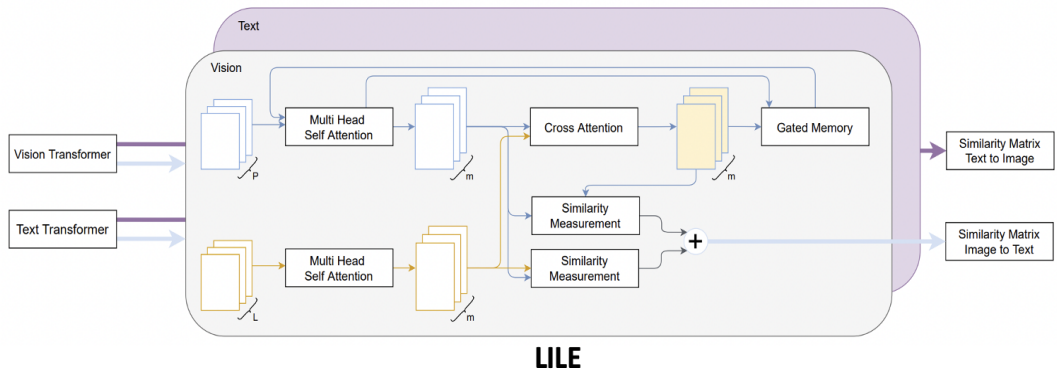

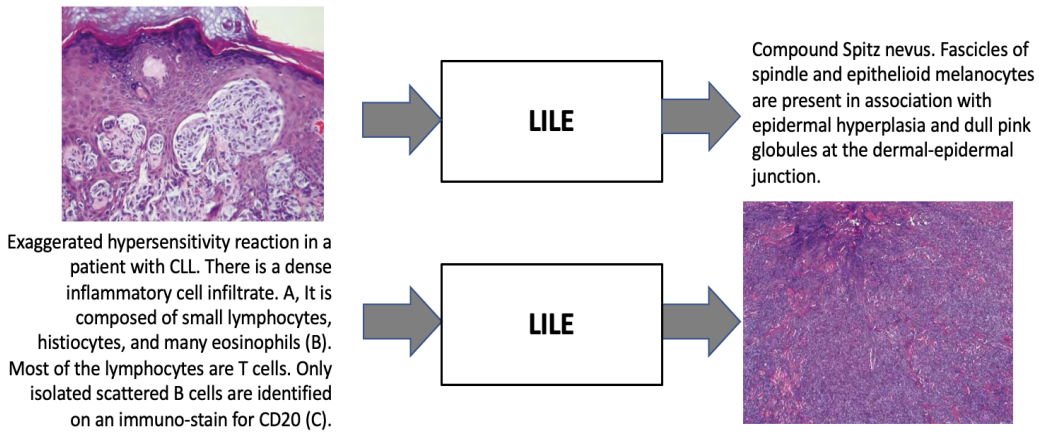

*Figure 6.* Bidirectional information retrieval. LILE (Look In-Depth before Looking Elsewhere) [Maleki2022] (top architects consisting of two transformers) uses a dual attention mechanism for cross-modal learning. This enables the model to auto-caption a query image (middle) as well as retrieve an image for a query description (bottom) (example inputs taken from ARCH [Gamper2021] processed by LILE to retrieve the output).

Major FMs include GPT-3 (multilingual translation, question answering, and creative text generation), Jurassic-1 Jumbo factual language processing), and LaMDA (generating realistic chat conversations). Generally, vision transformers can take large topologies to be used for object detection and image captioning. As a versatile base with vast knowledge learned from massive data, FMs can be finetuned for new downstream tasks and hence democratize AI usage by significantly reducing development time of deep approaches for a large number of applications.

Foundation models, as the superset of LLMs and LVLMs, may also be susceptible to biased responces and be unfair to specific social groups due to the inevitable societal preferences and predispositions in the training data [Dehkharghanian2023]. The "black-box" concern about neural networks becomes more pressing as the lack of explainability and interpretability becomes even more pronounced for how FMS make decisions. One must emphasize that the massive computational resources required for training, maintaining and running these models is certainly a major barrier for small clinics and community hospitals with limited resources. In general, the size of data and its relation to the size of a network can create different challenges. Whereas we have been mainly worried about overfitting, the advent of foundation models adds new concerns like 'hallucination' to our terminology (see Figure 8).

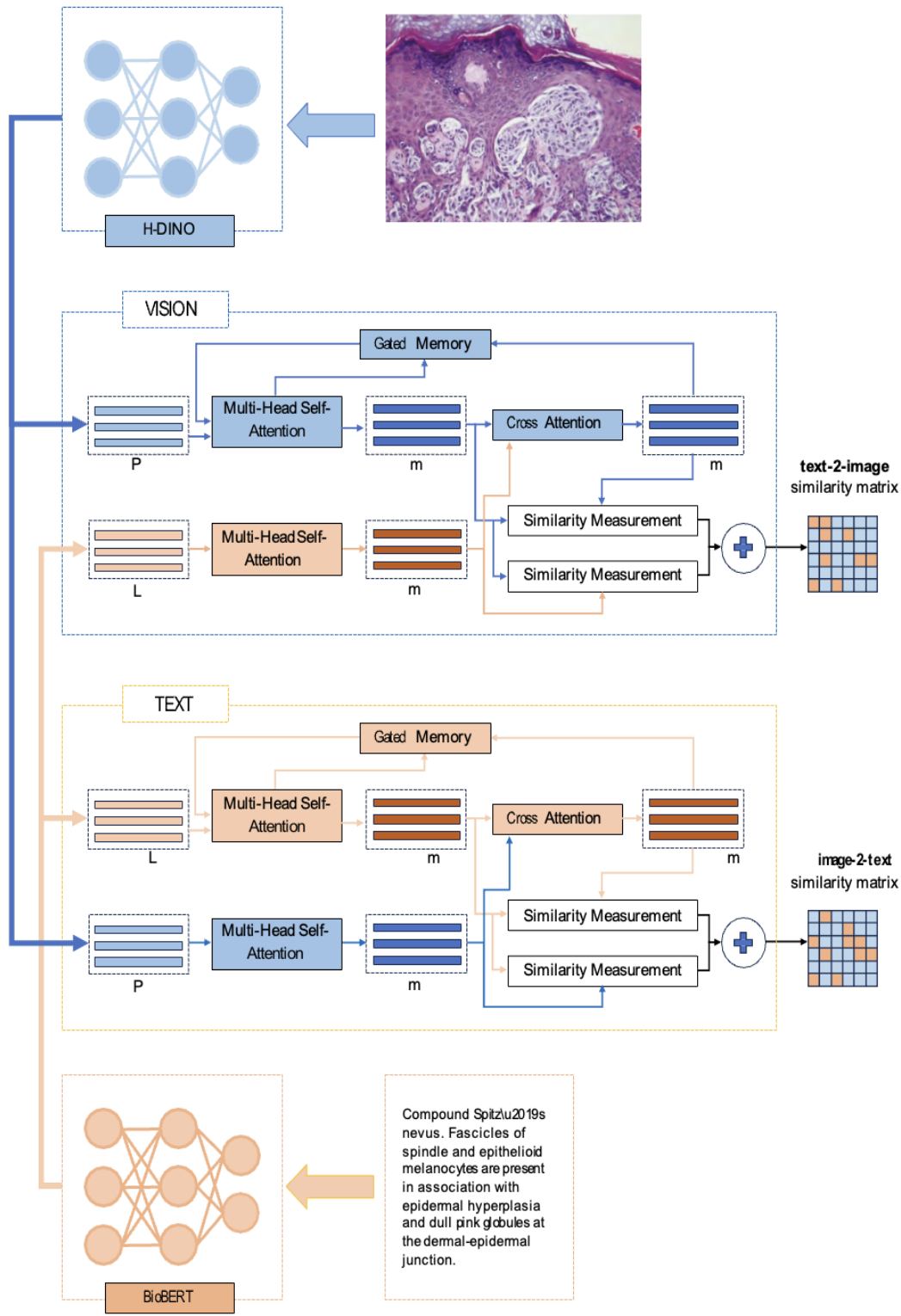

*Figure 7.* The architecture of harmonizing LILE using DINO for image-text data is shown. It comprises H-DINO architecture and BioBERT to extract features for both vision and text modalities and LILE backbone to align feature representation of pair images and text (Source: [Maleki2024]).

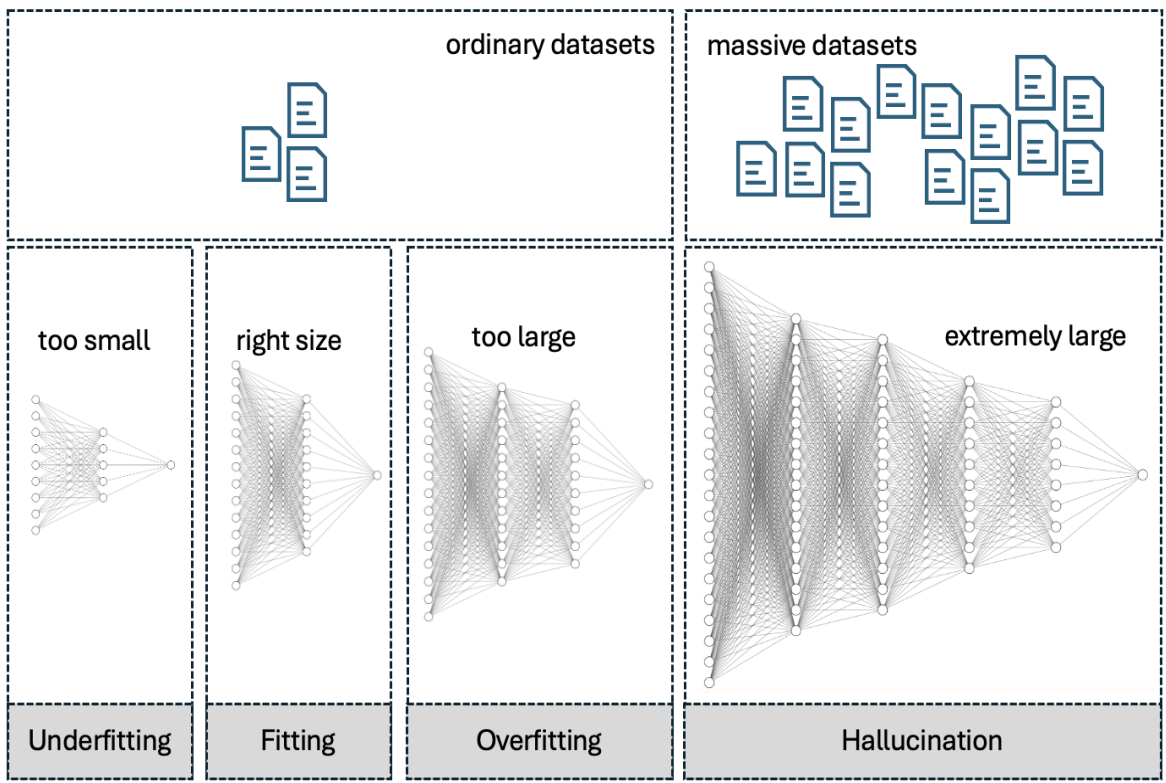

*Figure 8.* The relationship between size of data and size of neural networks may be optimal for fitting, or may cause different issues like under- and overfitting for conventional networks, and hallucinations for larger models (network depictions taken from alexlenail.me).

## Generative AI

Most AI models are discriminative in nature, meaning that they draw lines (or planes) to separate different classes against each other. A form of deep models that, instead of classification, can rather produce new data such as text and images are called Generative AI [Feuerriegel2024] (see Figure 9). Like any other AI approach, generative AI learns by being trained on available datasets, but rather on massive datasets in order to capture the data distribution and existing correlations among data instances. During such training, the data patterns are identified and processed to grasp the meaning of a cluster of pixels in an image or a group of words in a sentence [Epstein2023, Jo2023]. Furthermore, generative AI can 'manufacture' novel data (or a combination of data that has not been seen before). Once trained, the model applies its learned knowledge to produce new text/images, mincing the distribution and recreating the correlations learned during training. Additionally, generative AI can be adjusted, i.e., fine-tuned, through "prompting", allowing users to guide the type of content they desire [Dang2022].

Large models, and generally foundation models exhibit adaptability as a fundamental characteristic by handling various prompts and generating different new data formats (although they are supposed to be quite capable as they are as well, the co-called zero-shot learning). Generative models demonstrate versatile capabilities such as translating between many languages, generating digital images based on user descriptions, crafting various forms of text like poems, and answering questions at different levels of detail [Gozalo2023]. Table 4 summarizes the differences between generative versus discriminative models.

*Table 4. Discriminative versus generative AI.*

|  | **Discriminative AI** | **Generative AI** |
|---|---|---|
| Model size | Small (<100M) | Large (>300M) |
| Dominant learning mode | Supervised | Self-Supervised |
| Data size | Lots of data (e.g., 1M images) | Mass data (e.g., 400M images) |
| Task | Makes decisions | Creates content |
| Domain | Specific | General |
| Reliability | Consistent | May be distorted |

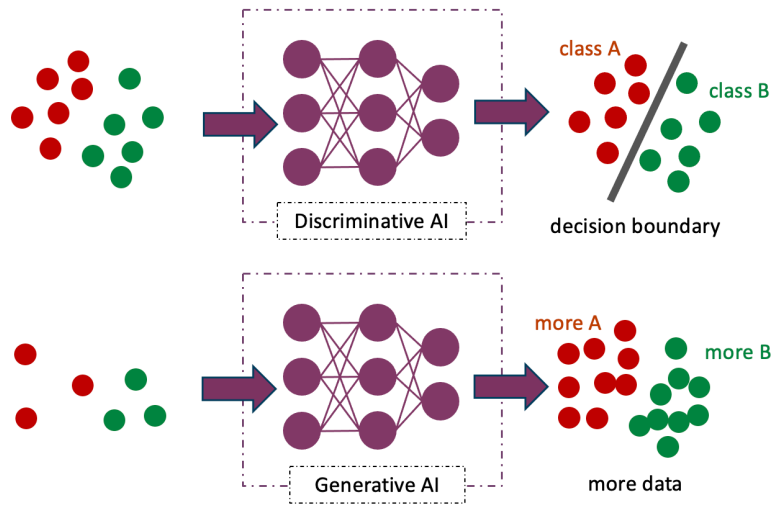

*Figure 9. The main task for discriminative AI is to classify data (top). Generative AI, in contrast, learns to generate more data of the same type/distribution.*

Various key topologies are employed to build generative AI models. One of the early and prominent approaches involves Generative Adversarial Networks (GANs) [Pan2019]. GANs mostly consist of two competitive parts: A *generator* produces (fake) data samples based on learned distribution, while a *discriminator* attempts to differentiate between real and fake samples. This 'adversarial' interaction guides the generator to generate better and better outputs to fool the discriminator, a framework that can even be used for generating complex patterns like human tissue [Safarpoor2021, Afshari2023]. Another generative scheme is implemented via Variational Autoencoders (VAEs) [Kingma2019]. VAEs also operate based on two main parts although quite different from GANs: an encoder network compresses data into a "core representation", and a decoder network reconstructs new data based on that core. VAEs can capture the saliencies and substance of training data, i.e., the complex nonlinear data distributions, and use it to generate similar examples. Lastly, transformer architectures [Vaswani2017, Hudson2021] have emerged as the most powerful generative engines. To accommodate data generation, they use many layers of purposefully arranged artificial neurons to analyze and connect different parts/fragments/segments of the input data. This exhaustive but soft correlation analysis, that can only be implemented in the form of a deep network, enables Transformers to understand correlations and complex relationships within the data and leverage this knowledge to generate new image/text content by activating trajectories of learned meaningful likelihoods.

# Information Retrieval and Foundation models

Considering the success and dominance of deep learning models, one may think that information retrieval may be replaced by foundation models. LLMs and LVLMs, and more generally foundation models, can perform conversations with the pathologists and, through displaying immense knowledge, convince the expert to accept what they are putting forward. However, the trend in developments seems to go toward the combination and merging of these two distinct technologies. Among possibilities, the Retrieval-Augmented Generation (RAG) is one of the major ideas that synergizes the capabilities of information retrieval and generative AI [Lewis202]. Among other, information retrieval can help foundation models with "source attribution" [Kamalloo2023] and point to the relevant information in small or large datasets solidifying the FM outputs with facts and details. This will reduce the "black box" disadvantage of models that persists in different forms also for FMs. On the other hand, generative AI excels at summarization, translation, and descriptions. However, FMs are limited in accessing the most accurate and up-to-date information and cannot point to the source.

Recently, some researchers have collected online images (e.g., from Twitter and PubMed) to retrain CLIP for histopathology [Huang2023, Lu2023]. The general wisdom in computer science is that employing online images is not a suitable venue: *Garbage in, garage out*. Validations with high-quality clinical data have clearly shown these models won't provide any value in pathology compared to simpler models trained with high-quality data [Alfasly2023, Alfasly2024]. Cross-modal retrieval with vision transformers (Figure 7), when trained with high-quality clinical data can indeed be much more reliable [Maleki2024]. In case the research community designs and trains a few foundation models for the pathology with high-quality data, one has to redefine the requirements for image search. In this case, the tasks will be intrinsically multimodal and the Divide & Conquer has been implemented within the pathology-aware foundation model. A new type of search will then perhaps be part of a FM-derived question and answering. Most crucial concern in this regard will be the "generative" aspect of FMs which is undesired in medicine when it degenerates into hallucinations. It's essential to note that foundation models with conversational capabilities are closely tied to search, serving as implicit information retrieval. Table 5 offers a comparison between conventional information retrieval and foundation models.

# Conclusions

With almost 400 years of history behind light microscopy, any innovation to bring about lasting change in histopathology may require some time. From today's perspective, deep learning's impact on digital (computational) pathology will be rather significant. The rise of large deep models is further amplifying the importance of deep networks in histopathology workflows. However, the future holds even greater promise when we regard the recent emergence of foundation models. These models have the potential to revolutionize computational pathology by offering valuable insights through "soft" information retrieval and fostering insightful conversations with researchers. While artificial intelligence through large foundational models, equipped with generative features of deep networks can excel at processing massive datasets and uncovering complex patterns, it cannot replace the need for traditional information retrieval platforms. As a matter of fact, AI is in dire need of information retrieval: Large or small, deep models struggle with transparency, simplicity, source attribution, and resource efficiency – all crucial aspects for their affordable and sustainable deployment in digital pathology. In contrast, conventional information retrieval methods like image search offer interpretable results, straightforward understanding, clear source attribution, and efficient resource utilization. Therefore, a comprehensive and balanced approach to information processing and knowledge extraction in digital pathology requires to focus on the synergy between large deep models and conventional information retrieval methods. This collaboration will leverage the strengths of both approaches to create more robust and informative systems of the future.

*Table 5.* Comparison of "search" versus "foundation models" as two technologies that may be used to assist pathologists in decision making (also see [Tizhoosh2024]).

| | **Information Retrieval (IR)** | **Foundation Models (FMs)** |
|---|---|---|
| **Model Size** | Small | Very Large |
| **Size of Dataset Needed** | Small | Very Large |
| **Computational Footprint** | Small | Very Large |
| **Strength** | IR convinces through retrieving evidence, i.e., the evidently diagnosed cases form the past | FMs convinces through knowledgeable conversations |
| **Disease Type Suitability** | All diseases including rare cases with only a few examples | Mainly common diseases with a lot of data available |
| **Information Processing Type** | Explicit Information Retrieval | Implicit Information Retrieval |
| **Source Attribution for responses to query** | Visible; Accessible; Explainable | Invisible; Not accessible; Not easily explainable |
| **Maintenance** | <ul><li>Low dependency on hardware updates</li><li>Adding/deleting cases straightforward</li><li>Newer models can replace the old ones</li></ul> | <ul><li>High dependency on hardware updates</li><li>High efforts for prompting to customize for specific tasks</li><li>Expensive re-training cycles may be necessary</li></ul> |

# References

[Afshari2023] Afshari, Mehdi, Saba Yasir, Gary L. Keeney, Rafael E. Jimenez, Joaquin J. Garcia, and Hamid R. Tizhoosh. "Single patch super-resolution of histopathology whole slide images: a comparative study." Journal of Medical Imaging 10, no. 1 (2023): 017501-017501.

[Alfasly2023] Alfasly, Saghir, Peyman Nejat, Sobhan Hemati, Jibran Khan, Isaiah Lahr, Areej Alsaafin, Abubakr Shafique et al. "When is a Foundation Model a Foundation Model." arXiv preprint arXiv:2309.11510 (2023).

[Alfasly2024] Alfasly, Saghir, Peyman Nejat, Sobhan Hemati, Jibran Khan, Isaiah Lahr, Areej Alsaafin, Abubakr Shafique et al. "Foundation Models for Histopathology—Fanfare or Flair." Mayo Clinic Proceedings: Digital Health 2, no. 1 (2024): 165-174.

[Alizadeh2023] Mohammad Alizadeh, Mohammad Sadegh Helfroush, and Henning Müller. A novel siamese deep hashing model for histopathology image retrieval. Expert Systems with Applications, 225:120169, 2023.

[Aljanabi2012] Al-Janabi, Shaimaa, André Huisman, and Paul J. Van Diest. "Digital pathology: current status and future perspectives." Histopathology 61, no. 1 (2012): 1-9.

[Arvisais2024] Arvisais-Anhalt, Simone, Steven L. Gonias, and Sara G. Murray. "Establishing priorities for implementation of large language models in pathology and laboratory medicine." Academic Pathology 11, no. 1 (2024): 100101.

[Babaie2017] Babaie, Morteza, Shivam Kalra, Aditya Sriram, Christopher Mitcheltree, Shujin Zhu, Amin Khatami, Shahryar Rahnamayan, and Hamid R. Tizhoosh. "Classification and retrieval of digital pathology scans: A new dataset." In Proceedings of the IEEE conference on computer vision and pattern recognition workshops, pp. 8-16. 2017.

[Bera2019] Bera, Kaustav, Kurt A. Schalper, David L. Rimm, Vamsidhar Velcheti, and Anant Madabhushi. "Artificial intelligence in digital pathology—new tools for diagnosis and precision oncology." Nature reviews Clinical oncology 16, no. 11 (2019): 703-715.

[Bommasani2021] Bommasani, Rishi, Drew A. Hudson, Ehsan Adeli, Russ Altman, Simran Arora, Sydney von Arx, Michael S. Bernstein et al. "On the opportunities and risks of foundation models." arXiv preprint arXiv:2108.07258 (2021).

[Browning2021] Browning, Lisa, Richard Colling, Emad Rakha, Nasir Rajpoot, Jens Rittscher, Jacqueline A. James, Manuel Salto-Tellez, David RJ Snead, and Clare Verrill. "Digital pathology and artificial intelligence will be key to supporting clinical and academic cellular pathology through COVID-19 and future crises: the PathLAKE consortium perspective." Journal of clinical pathology 74, no. 7 (2021): 443-447.

[Caron2021] Caron, Mathilde, Hugo Touvron, Ishan Misra, Hervé Jégou, Julien Mairal, Piotr Bojanowski, and Armand Joulin. "Emerging properties in self-supervised vision transformers." In Proceedings of the IEEE/CVF international conference on computer vision, pp. 9650-9660. 2021.

[Chaudhary2016] Chaudhary, Kamal Kumar, and Nidhi Mishra. "A review on molecular docking: novel tool for drug discovery." Databases 3, no. 4 (2016): 1029.

[Chen2022] Chengkuan Chen, Ming Y. Lu, Drew F. K. Williamson, Tiffany Y. Chen, Andrew J. Schaumberg, and Faisal Mahmood. Fast and scalable search of whole-slide images via self-supervised deep learning. NatureBiomedical Engineering, 6(12):1420–1434, 2022.

[Choi2023] Choi, Hyeon Seok, Jun Yeong Song, Kyung Hwan Shin, Ji Hyun Chang, and Bum-Sup Jang. "Developing prompts from large language model for extracting clinical information from pathology and ultrasound reports in breast cancer." Radiation Oncology Journal 41, no. 3 (2023): 209.

[Cornsweet2012] Cornsweet, Tom. Visual perception. Academic press, 2012.

[Dang2022] Dang, Hai, Lukas Mecke, Florian Lehmann, Sven Goller, and Daniel Buschek. "How to prompt? Opportunities and challenges of zero-and few-shot learning for human-AI interaction in creative applications of generative models." arXiv preprint arXiv:2209.01390 (2022).

[Dehkharghanian2023] Dehkharghanian, Taher, Azam Asilian Bidgoli, Abtin Riasatian, Pooria Mazaheri, Clinton JV Campbell, Liron Pantanowitz, H. R. Tizhoosh, and Shahryar Rahnamayan. "Biased data, biased AI: deep networks predict the acquisition site of TCGA images." Diagnostic pathology 18, no. 1 (2023): 67.

[Epstein2023] Epstein, Ziv, Aaron Hertzmann, Investigators of Human Creativity, Memo Akten, Hany Farid, Jessica Fjeld, Morgan R. Frank et al. "Art and the science of generative AI." Science 380, no. 6650 (2023): 1110-1111.

[Evans2022] Evans, Andrew J., Richard W. Brown, Marilyn M. Bui, Elizabeth A. Chlipala, Christina Lacchetti, Danny A. Milner Jr, Liron Pantanowitz et al. "Validating whole slide imaging systems for diagnostic purposes in pathology: guideline update from the College of American Pathologists in collaboration with the American Society for Clinical Pathology and the Association for Pathology Informatics." Archives of pathology & laboratory medicine 146, no. 4 (2022): 440-450.

[Feuerriegel2024] Feuerriegel, Stefan, Jochen Hartmann, Christian Janiesch, and Patrick Zschech. "Generative ai." Business & Information Systems Engineering 66, no. 1 (2024): 111-126.

[Freire2024] Freire, Samuel Kernan, Chaofan Wang, and Evangelos Niforatos. "Chatbots in Knowledge-Intensive Contexts: Comparing Intent and LLM-Based Systems." arXiv preprint arXiv:2402.04955 (2024).

[Gamper2021] Gamper, Jevgenij, and Nasir Rajpoot. "Multiple instance captioning: Learning representations from histopathology textbooks and articles." In Proceedings of the IEEE/CVF conference on computer vision and pattern recognition, pp. 16549-16559. 2021.

[Gozalo2023] Gozalo-Brizuela, Roberto, and Eduardo C. Garrido-Merchán. "A survey of Generative AI Applications." arXiv preprint arXiv:2306.02781 (2023).

[Griffin2017] Griffin, Jon, and Darren Treanor. "Digital pathology in clinical use: where are we now and what is holding us back?." Histopathology 70, no. 1 (2017): 134-145.

[Guo2016] Guo, Jiafeng, Yixing Fan, Qingyao Ai, and W. Bruce Croft. "Semantic matching by non-linear word transportation for information retrieval." In Proceedings of the 25th ACM International on Conference on Information and Knowledge Management, pp. 701-710. 2016.

[Hadi2023] Hadi, Muhammad Usman, Rizwan Qureshi, Abbas Shah, Muhammad Irfan, Anas Zafar, Muhammad Bilal Shaikh, Naveed Akhtar, Jia Wu, and Seyedali Mirjalili. "A survey on large language models: Applications, challenges, limitations, and practical usage." Authorea Preprints (2023).

[Hanna2020] Hanna, Matthew G., Victor E. Reuter, Orly Ardon, David Kim, Sahussapont Joseph Sirintrapun, Peter J. Schüffler, Klaus J. Busam et al. "Validation of a digital pathology system including remote review during the COVID-19 pandemic." Modern Pathology 33, no. 11 (2020): 2115-2127.

[Hanna2022] Hanna, Matthew G., Orly Ardon, Victor E. Reuter, Sahussapont Joseph Sirintrapun, Christine England, David S. Klimstra, and Meera R. Hameed. "Integrating digital pathology into clinical practice." Modern Pathology 35, no. 2 (2022): 152-164.

[Hart2023] Hart, Steven N., Noah G. Hoffman, Peter Gershkovich, Chancey Christenson, David S. McClintock, Lauren J. Miller, Ronald Jackups, Vahid Azimi, Nicholas Spies, and Victor Brodsky. "Organizational preparedness for the use of large language models in pathology informatics." Journal of Pathology Informatics (2023): 100338.

[Hegde2019] Narayan Hegde, Jason D Hipp, Yun Liu, Michael Emmert-Buck, Emily Reif, Daniel Smilkov, Michael Terry, Carrie J Cai, Mahul B Amin, Craig H Mermel, et al. Similar image search for histopathology: Smily.NPJ digital medicine, 2(1):56, 2019.

[Hersh2008] Hersh, William. Information retrieval: a health and biomedical perspective. Springer Science & Business Media, 2008.

[Huang2023] Huang, Zhi, Federico Bianchi, Mert Yuksekgonul, Thomas J. Montine, and James Zou. "A visual–language foundation model for pathology image analysis using medical Twitter." Nature Medicine (2023): 1-10.

[Hudson2021] Hudson, Drew A., and Larry Zitnick. "Generative adversarial transformers." In International conference on machine learning, pp. 4487-4499. PMLR, 2021.

[Jo2023] Jo, A. "The promise and peril of generative AI." Nature 614, no. 1 (2023): 214-216.

[Kalra2020a] Kalra Shivam, H. R. Tizhoosh, Choi Charles, Shah Sultaan, Diamandis Phedias, J. V. Campbell Clinton, and Pantanowitz Liron. Yottixel – an image search engine for large archives of histopathology whole slide images. Medical Image Analysis, 65:101757, 2020.

[Kalra2020b] Kalra, Shivam, Hamid R. Tizhoosh, Sultaan Shah, Charles Choi, Savvas Damaskinos, Amir Safarpoor, Sobhan Shafiei et al. "Pan-cancer diagnostic consensus through searching archival histopathology images using artificial intelligence." NPJ digital medicine 3, no. 1 (2020): 31.

[Kamalloo2023] Kamalloo, Ehsan, Aref Jafari, Xinyu Zhang, Nandan Thakur, and Jimmy Lin. "Hagrid: A human-llm collaborative dataset for generative information-seeking with attribution." arXiv preprint arXiv:2307.16883 (2023).

[Kamath2021] Kamath, Sowmya, Veena Mayya, and R. Priyadarshini. "A Probabilistic Precision Information Retrieval Model for Personalized Clinical Trial Recommendation based on Heterogeneous Data." In 2021 12th International Conference on Computing Communication and Networking Technologies (ICCCNT), pp. 1-5. IEEE, 2021.

[Kingma2019] Kingma, Diederik P., and Max Welling. "An introduction to variational autoencoders." Foundations and Trends® in Machine Learning 12, no. 4 (2019): 307-392.

[Kiran2023] Kiran, Nfn, F. N. U. Sapna, F. N. U. Kiran, Deepak Kumar, F. N. U. Raja, Sheena Shiwlani, Antonella Paladini et al. "Digital Pathology: Transforming Diagnosis in the Digital Age." Cureus 15, no. 9 (2023).

[Klein2021] Klein, Christophe, Qinghe Zeng, Floriane Arbaretaz, Estelle Devêvre, Julien Calderaro, Nicolas Lomenie, and Maria Chiara Maiuri. "Artificial intelligence for solid tumour diagnosis in digital pathology." British Journal of Pharmacology 178, no. 21 (2021): 4291-4315.

[Kumar2020] Kumar, Neeta, Ruchika Gupta, and Sanjay Gupta. "Whole slide imaging (WSI) in pathology: current perspectives and future directions." Journal of digital imaging 33, no. 4 (2020): 1034-1040.

[Laban2023] Laban, Philippe, Wojciech Kryściński, Divyansh Agarwal, Alexander Richard Fabbri, Caiming Xiong, Shafiq Joty, and Chien-Sheng Wu. "SummEdits: Measuring LLM ability at factual reasoning through the lens of summarization." In Proceedings of the 2023 Conference on Empirical Methods in Natural Language Processing, pp. 9662-9676. 2023.

[Lahr2024] Lahr, Isaiah, Saghir Alfasly, Peyman Nejat, Jibran Khan, Luke Kottom, Vaishnavi Kumbhar, Areej Alsaafin et al. "Analysis and Validation of Image Search Engines in Histopathology." arXiv preprint arXiv:2401.03271 (2024).

[Laurianne2020] David, Laurianne, Amol Thakkar, Rocío Mercado, and Ola Engkvist. "Molecular representations in AI-driven drug discovery: a review and practical guide." Journal of Cheminformatics 12, no. 1 (2020): 1-22.

[Lewis202] Lewis, Patrick, Ethan Perez, Aleksandra Piktus, Fabio Petroni, Vladimir Karpukhin, Naman Goyal, Heinrich Küttler et al. "Retrieval-augmented generation for knowledge-intensive nlp tasks." Advances in Neural Information Processing Systems 33 (2020): 9459-9474.

[Li2023] Shengrui Li, Yining Zhao, Jun Zhang, Ting Yu, Ji Zhang, and Yue Gao.High-order correlation-guided slide-level histology retrieval with self-supervised hashing. IEEE Transactions on Pattern Analysis and Machine Intelligence, 2023.

[Liu2024] Liu, Yiheng, Hao He, Tianle Han, Xu Zhang, Mengyuan Liu, Jiaming Tian, Yutong Zhang et al. "Understanding llms: A comprehensive overview from training to inference." arXiv preprint arXiv:2401.02038 (2024).

[Lu2023] Lu, Ming Y., Bowen Chen, Drew FK Williamson, Richard J. Chen, Ivy Liang, Tong Ding, Guillaume Jaume et al. "Towards a Visual-Language Foundation Model for Computational Pathology." arXiv preprint arXiv:2307.12914 (2023).

[Maleki2022] Maleki, D., & Tizhoosh, H. R. (2022, December). LILE: Look in-depth before looking elsewhere–a dual attention network using transformers for cross-modal information retrieval in histopathology archives. In International Conference on Medical Imaging with Deep Learning (pp. 879-894). PMLR.

[Maleki2024] Danial Maleki, Shahryar Rahnamayan, and H.R Tizhoosh, A Self-Supervised Framework for Cross-Modal Search in Histopathology Archives Using Scale Harmonization. Preprint on Springer's Research Square, 2024.

[Manning2009] Manning, Christopher D. An introduction to information retrieval. Cambridge university press, 2009.

[McInerney2020] McInerney, Denis Jered, Borna Dabiri, Anne-Sophie Touret, Geoffrey Young, Jan-Willem Meent, and Byron C. Wallace. "Query-focused ehr summarization to aid imaging diagnosis." In Machine Learning for Healthcare Conference, pp. 632-659. PMLR, 2020.

[Niazi2019] Niazi, Muhammad Khalid Khan, Anil V. Parwani, and Metin N. Gurcan. "Digital pathology and artificial intelligence." The lancet oncology 20, no. 5 (2019): e253-e261.

[Pan2019] Pan, Zhaoqing, Weijie Yu, Xiaokai Yi, Asifullah Khan, Feng Yuan, and Yuhui Zheng. "Recent progress on generative adversarial networks (GANs): A survey." IEEE access 7 (2019): 36322-3633

[Pantanowitz2010] Pantanowitz, Liron. "Digital images and the future of digital pathology." Journal of pathology informatics 1 (2010).

[Pantanowitz2021] Pantanowitz, Liron, Pamela Michelow, Scott Hazelhurst, Shivam Kalra, Charles Choi, Sultaan Shah, Morteza Babaie, and Hamid R. Tizhoosh. "A digital pathology solution to resolve the tissue floater conundrum." Archives of pathology & laboratory medicine 145, no. 3 (2021): 359-364.

[Parwani2020] Parwani, Anil V., and Mahul B. Amin. "Convergence of digital pathology and artificial intelligence tools in anatomic pathology practice: current landscape and future directions." Advances in Anatomic Pathology 27, no. 4 (2020): 221-226.

[Peris2023] Peris, Charith, Christophe Dupuy, Jimit Majmudar, Rahil Parikh, Sami Smaili, Richard Zemel, and Rahul Gupta. "Privacy in the Time of Language Models." In Proceedings of the Sixteenth ACM International Conference on Web Search and Data Mining, pp. 1291-1292. 2023.

[Radford2021] Radford, Alec, Jong Wook Kim, Chris Hallacy, Aditya Ramesh, Gabriel Goh, Sandhini Agarwal, Girish Sastry et al. "Learning transferable visual models from natural language supervision." In International conference on machine learning, pp. 8748-8763. PMLR, 2021.

[Rawte2023] Rawte, Vipula, Amit Sheth, and Amitava Das. "A survey of hallucination in large foundation models." arXiv preprint arXiv:2309.05922 (2023).

[Reis2016] Dos Reis, Julio Cesar, Rodrigo Bonacin, and Edemar Mendes Perciani. "Intention-based information retrieval of electronic health records." In 2016 IEEE 25th International Conference on Enabling Technologies: Infrastructure for Collaborative Enterprises (WETICE), pp. 217-222. IEEE, 2016.

[Rosai2007] Rosai, Juan. "Why microscopy will remain a cornerstone of surgical pathology." Laboratory investigation 87, no. 5 (2007): 403-408.

[Safarpoor2021] Safarpoor, Amir, Shivam Kalra, and Hamid R. Tizhoosh. "Generative models in pathology: synthesis of diagnostic quality pathology images." The Journal of Pathology 253, no. 2 (2021): 131-132.

[Salewski2024] Salewski, Leonard, Stephan Alaniz, Isabel Rio-Torto, Eric Schulz, and Zeynep Akata. "In-Context Impersonation Reveals Large Language Models' Strengths and Biases." Advances in Neural Information Processing Systems 36 (2024).

[Sikaroudi2023] Sikaroudi, Milad, Mehdi Afshari, Abubakr Shafique, Shivam Kalra, and Hamid R. Tizhoosh. "Comments on'Fast and scalable search of whole-slide images via self-supervised deep learning'." arXiv preprint arXiv:2304.08297 (2023).

[Singhal2001] Singhal, Amit. "Modern information retrieval: A brief overview." IEEE Data Eng. Bull. 24, no. 4 (2001): 35-43.

[Sivarajkumar2024] Sivarajkumar, Sonish, Haneef Ahamed Mohammad, David Oniani, Kirk Roberts, William Hersh, Hongfang Liu, Daqing He, Shyam Visweswaran, and Yanshan Wang. "Clinical information retrieval: A literature review." Journal of Healthcare Informatics Research (2024): 1-40.

[Suetens2017] Suetens, Paul. Fundamentals of medical imaging. Cambridge university press, 2017.

[Tayebi2022] Tayebi, R. Moosavi, Youqing Mu, Taher Dehkharghanian, Catherine Ross, Monalisa Sur, Ronan Foley, Hamid R. Tizhoosh, and Clinton JV Campbell. "Automated bone marrow cytology using deep learning to generate a histogram of cell types." *Communications medicine* 2, no. 1 (2022): 45.

[Ting2013] Ting, S. L., Eric WK See-To, and Ying Kei Tse. "Web information retrieval for health professionals." Journal of medical systems 37 (2013): 1-14.

[Tizhoosh2018] Tizhoosh, Hamid Reza, and Liron Pantanowitz. "Artificial intelligence and digital pathology: challenges and opportunities." Journal of pathology informatics 9, no. 1 (2018): 38.

[Tizhoosh2021] Tizhoosh, Hamid R., Phedias Diamandis, Clinton JV Campbell, Amir Safarpoor, Shivam Kalra, Danial Maleki, Abtin Riasatian, and Morteza Babaie. "Searching images for consensus: can AI remove observer variability in pathology?." The American journal of pathology 191, no. 10 (2021): 1702-1708.

[Tizhoosh2022] Tizhoosh, Hamid Reza. "Systems and methods for barcode annotations for digital images." U.S. Patent 11,270,204, issued March 8, 2022.

[Tizhoosh2024] Tizhoosh, H. R., and Liron Pantanowitz. "On Image Search in Histopathology." arXiv preprint arXiv:2401.08699 (2024).

[Touvron2023] Touvron, Hugo, Louis Martin, Kevin Stone, Peter Albert, Amjad Almahairi, Yasmine Babaei, Nikolay Bashlykov et al. "Llama 2: Open foundation and fine-tuned chat models." arXiv preprint arXiv:2307.09288 (2023).

[Tseng2023] Tseng, Liang-Jun, Arata Matsuyama, and Valerie MacDonald-Dickinson. "Histology: The gold standard for diagnosis?." The Canadian Veterinary Journal 64, no. 4 (2023): 389.

[Ullah2024] Ullah, Ehsan, Anil Parwani, Mirza Mansoor Baig, and Rajendra Singh. "Challenges and barriers of using large language models (LLM) such as ChatGPT for diagnostic medicine with a focus on digital pathology–a recent scoping review." Diagnostic Pathology 19, no. 1 (2024): 1-9.

[Vanopstal2013] Vanopstal, Klaar, Joost Buysschaert, Godelieve Laureys, and Robert Vander Stichele. "Lost in PubMed. Factors influencing the success of medical information retrieval." Expert Systems with Applications 40, no. 10 (2013): 4106-4114.

[Vaswani2017] Vaswani, Ashish, Noam Shazeer, Niki Parmar, Jakob Uszkoreit, Llion Jones, Aidan N. Gomez, Łukasz Kaiser, and Illia Polosukhin. "Attention is all you need." Advances in neural information processing systems 30 (2017).

[Wang2019] Wang, Yanshan, Andrew Wen, Sijia Liu, William Hersh, Steven Bedrick, and Hongfang Liu. "Test collections for electronic health record-based clinical information retrieval." JAMIA open 2, no. 3 (2019): 360-368.

[Wang2023] Xiyue Wang, Yuexi Du, Sen Yang, Jun Zhang, Minghui Wang, Jing Zhang, Wei Yang, Junzhou Huang, and Xiao Han. Retccl: clustering-guided contrastive learning for whole-slide image retrieval. Medical image analysis, 83:102645, 2023

[Wu2023] Wu, Tianyu, Shizhu He, Jingping Liu, Siqi Sun, Kang Liu, Qing-Long Han, and Yang Tang. "A brief overview of ChatGPT: The history, status quo and potential future development." IEEE/CAA Journal of Automatica Sinica 10, no. 5 (2023): 1122-1136.

[Xu2024] Xu, Ziwei, Sanjay Jain, and Mohan Kankanhalli. "Hallucination is inevitable: An innate limitation of large language models." arXiv preprint arXiv:2401.11817 (2024).

[Yang2023] Yang, Jingfeng, Hongye Jin, Ruixiang Tang, Xiaotian Han, Qizhang Feng, Haoming Jiang, Shaochen Zhong, Bing Yin, and Xia Hu. "Harnessing the power of llms in practice: A survey on chatgpt and beyond." ACM Transactions on Knowledge Discovery from Data (2023).

[Zarella2023] Zarella, Mark D., David S. McClintock, Harsh Batra, Rama R. Gullapalli, Michael Valante, Vivian O. Tan, Shubham Dayal et al. "Artificial intelligence and digital pathology: clinical promise and deployment considerations." Journal of Medical Imaging 10, no. 5 (2023): 051802-051802.

[Zhang2014] Zhang, Xiaofan, Wei Liu, Murat Dundar, Sunil Badve, and Shaoting Zhang. "Towards large-scale histopathological image analysis: Hashing-based image retrieval." IEEE Transactions on Medical Imaging 34, no. 2 (2014): 496-506.

[Zhang2023] Zhang, Zheng, Chen Zheng, Da Tang, Ke Sun, Yukun Ma, Yingtong Bu, Xun Zhou, and Liang Zhao. "Balancing specialized and general skills in llms: The impact of modern tuning and data strategy." arXiv preprint arXiv:2310.04945 (2023).

[Zhao2023] Zhao, Theodore, Mu Wei, J. Samuel Preston, and Hoifung Poon. "Automatic calibration and error correction for large language models via pareto optimal self-supervision." arXiv preprint arXiv:2306.16564 (2023).

[Zhou2008] Zhou, Xiang Sean, Sonja Zillner, Manuel Moeller, Michael Sintek, Yiqiang Zhan, Arun Krishnan, and Alok Gupta. "Semantics and CBIR: a medical imaging perspective." In Proceedings of the 2008 international conference on Content-based image and video retrieval, pp. 571-580. 2008.

[Zhu2018] Shujin Zhu, Yuehua Li, Shivam Kalra, and Hamid R Tizhoosh. Multiple disjoint dictionaries for representation of histopathology images. Journal of Visual Communication and Image Representation, 55:243–252, 2018.

{Zhao2023] Zhao, Wayne Xin, Kun Zhou, Junyi Li, Tianyi Tang, Xiaolei Wang, Yupeng Hou, Yingqian Min et al. "A survey of large language models." arXiv preprint arXiv:2303.18223 (2023).